\documentclass[showpacs,amsmath,amssymb]{revtex4-2}

\usepackage{graphicx}
\newcommand{\be}{\begin{equation}}
\newcommand{\ee}{\end{equation}}
\newcommand{\ba}{\begin{eqnarray}}
\newcommand{\ea}{\end{eqnarray}}

\newcommand{\beq}{\begin{equation}}
\newcommand{\eeq}{\end{equation}}
\newcommand{\beqs}{\begin{eqnarray}}
\newcommand{\eeqs}{\end{eqnarray}}

\newfont{\prg}{cmsy10}

\begin{document}
\title{ 
 New features in the differential cross sections  measured at the LHC
   }
\author{
O.~V. Selyugin \thanks{selugin@thsun1.jinr.ru}}

\affiliation
{  
 \it BLTP,
Joint Institute for Nuclear Research,
141980 Dubna, Moscow region, Russia}

\begin{abstract}
 The critical  analysis of the new experimental data obtained by
 the ATLAS Collaboration group at 13 TeV  is presented and the problem of the tension
 between data of the ATLAS and TOTEM Collaboration is considered.
 The analysis of  new effects discovered on the basis of experimental  data at 13 TeV \cite{osc13,fd13}
  and associated with the specific properties of the  hadron potential
  at large distances is carried out
   taking account of all sets of experimental data on elastic $pp$-scattering obtained by  the TOTEM and ATLAS Collaborations in a wide momentum transfer region.
   It also  gives quantitative descriptions of all examined experimental data
    with a minimum of fitting parameters.
   It is shown that the new  features determined at a  high statistical level
   give an important contribution to the differential cross sections and allow
    the research into hadron interactions at large distances.
   \end{abstract}

 \pacs{11.80.Cr, 12.40.Nn, 13.85.Dz}
 \maketitle


 \section{Introduction}

    To get hard predictions for hadronic interactions at ultrahigh energies, we need
    to know with the highest possible accuracy the structure of hadrons and their interactions.
   The main purpose of a new accelerator is to find  new effects in  particle interactions.
  This especially concerns studies of elastic hadron scattering at the LHC.
   It gives   information about  hadron interaction
   at large distances where  perturbative QCD does not work
   and a new theory such as instanton or string theory,
    has to be developed.

		 The first measurements of the TOTEM Collaboration  confirmed  the non-linear
   behavior of the slope of the differential cross sections at a small momentum transfer \cite{T8nexp}, 
   which shows a complicated picture of the hadron interaction at super high energies.
  The unique experiments of the TOTEM-SMS and ALFA-ATLAS Collaborations at 13 TeV gave experimental data
   with unprecedentedly  small  scattering angles  and narrow $\Delta t$ at small momentum transfer \cite{T67,ATLAS-13}.
   However, it is necessary to note some problems with the normalization of experimental data.

     The study of the experimental data at $13$ TeV obtained by the TOTEM Collaboration
     reveals new effects in the behavior of differential cross sections, oscillations \cite{osc13},
     which are practically   independent of the normalization of differential cross sections
    (see below),
     as well as an additional anomalous term
      in the scattering amplitude with a large slope \cite{fd13}, which gives 
       necessary contributions at small momentum transfer.
     It is important to study  the full sets of experimental data obtained at LHC $7 \leq \sqrt{s} \leq 13$ TeV by both
     Collaborations. For this purpose,   the high energy generalized structure (HEGS) model was chosen as it has a
     minimum fitting parameters and gives a quantitative description of the differential cross section
     in a huge energy region   $9 \leq \sqrt{s} \leq 13 000$ GeV and simultaneously in the Coulomb-hadron interference region
     and a large momentum transfer \cite{HEGS0,HEGS1,HEGSdm}.

       In the second section of the paper,    a short analysis of experimental data on the differential cross sections of elastic
       proton-proton scattering is presented. In the third section, the new effects are analyzed in the framework of the HEGS model
   taking into account  experimental data of all sets of the TOTEM and ATLAS Collaborations
   obtained at $2.7 -13 $ TeV.
 In the fourth sections, the results and some consequences are discussed.
 The conclusions are given in the final section.

\section{Experimental data }

Now there are 9 independent sets of  elastic $pp$ differential cross sections
obtained by the ATLAS and TOTEM Collaborations at LHC energies $\sqrt{s} = 7 - 13 $ TeV,
6 of them are measured at a small momentum transfer ($t$)
including the Coulomb-hadron interference  region
(2 obtained by ATLAS and 3 by the TOTEM Collaboration). The other 3 sets are measured at large $t$
including the region of the dip-bump structure.
At small $t$
the values of $\sigma_{tot}(s)$ and $\rho(s,t=0)$ were obtained
on the basis of experimental data.
Some tension between $\sigma_{tot}(s)$ obtained by the ATLAS and TOTEM Collaborations  should be noted.
The difference is $3.15$ mb at $7$ TeV and grows to $5.6$ mb at $8$ TeV.
Note that in \cite{Panch}  some contradictions between the obtained $\sigma_{tot}(s)$ and $\rho(s,t=0)$
by the TOTEM at $7$ TeV were shown.
In the paper \cite{D-L-13},  some discrepancy
between the slopes of the differential cross sections measured by the TOTEM at $7$ and $8$ TeV were noted.

Until recently,
we had no  ATLAS-AlFA data at 13 TeV.
Now these data are published \cite{ATLAS-13}. The tension between values of $\sigma_{tot}(s)$
increases up to $5.75$ mb.
In a recent paper \cite{Inel13}, the importance of the contribution of inelastic
scattering at small angles, especially in the case of
determining  the total cross section  by the luminosity independent method.
We can remember the large difference  in the measurements
of  $\sigma_{tot}(s)$  by the luminosity independent method ($72$ and $80$ mb) by  two Collaborations at the Tevatron.
However, in our opinion the situation is more complicated
and   requires  careful analysis of the new data.

\begin{table}
\label{Table_1}
\caption{ Sets of $d\sigma/dt$ and $\chi^{2}_{dof}$ obtained in the HEGS model
($k_i$ -additional normalization coefficient)  }
\vspace{.5cm}
\begin{tabular}{|c|c|c|c|c|c|c|c|} \hline
 $\sqrt{s}$ & $n$ &$-t_{min}$ & $-t_{max}$ & $\sum \chi^2_{i}$ & $\chi_{dof}$  &$k_{norm}$     \\ TeV        & $n$ &$(GeV^2)$  &$(GeV^2)$   &                   &               &    \\ \hline
 7  T(1)  &79    & 0.0699 & 162.1& 2.19 &$1.12  $ & $ 1.01    $ \\
 7  A(1)  &70    & 0.0559 & 86.5 & 1.29 &$1.15  $ & $ 1.04   $ \\
 8  T(2)  &65    & 0.0488 & 66.3 & 1.07 &$1.16  $ & $ 0.93   $ \\
 8  A(2)  &60    & 0.0422 & 55.3 & 0.97 &$1.16  $ & $ 1.04   $ \\
 8  T(3)  &55    & 0.0361 & 49.7 & 0.96 &$1.16  $ & $ 1.04   $ \\
 8  T(4)  &50    & 0.0305 & 47.8 & 1.02 &$1.01  $ & $ 0.94   $ \\
13  T(5)  &40    & 0.0207 & 34.2 & 0.92 &$1.11  $ & $ 1.06   $ \\       \hline
$\sum$  AT &712    & 0.0007 & 2.4 & 715 &$ 1.01  $ & $\overline{1.06} $   \\       \hline
\end{tabular}
\end{table}
\vspace{.5cm}

  The new experiment of  the ATLAS-ALFA Collaboration \cite{ATLAS-13} presented
  unique results.
  The position of Roman Pot situated at very large distances (2400 m) allows one to obtain
   new data at very small angles with small $\Delta t$. The experiment  for first time gave five  new  experimental points in the essential Coulomb-nuclear interference region.
   The first experimental point
  begins from $-t=0.00025$ GeV$^2$ with $\Delta t = 0.0001$ GeV$^2$.
   To our regret, this first point has a large error (only statistical error reaching $26$ per cent).
   However, there are  at least six-seven more points in the Coulomb Nuclear Interference(CNI) region. Three-four of them
   overlap the region where the TOTEM Collaboration gave own experimental points.
   Obviously, we can see that the  ATLAS points  lie systematically below the TOTEM  points.
   Hence, they require some additional  normalization. This is so,  and our model description of both the sets
   shows, in all variants of the model, the difference in the normalization is about $11\% - 13\%$.
   As the points lie in the 
   significant CNI region, we can check up these experimental points of ATLAS
   by comparing their  momentum transfer dependence with the $t$ dependence
   of the contribution of the Coulomb amplitude.

     Let us extract from the differential cross section of  elastic scattering the term of  Coulomb-nuclear interference,  which is determined by
     the contribution of the pure Coulomb amplitude and the contribution of the pure hadron amplitude
    \ba
 \frac{d\sigma}{dt}|_{CNI}=  \frac{d\sigma}{dt}|_{exper.}- \frac{d\sigma}{dt}|_{F_{C}^{2}}- \frac{d\sigma}{dt}|_{Fh^{2}}
\ea

\vspace{3.5cm}
  \begin{table}
\label{Table-2}
\caption{  $\Delta_{CNI}=d\sigma/dt_{CNI}$ mb/GeV$^{2}$   }
\vspace{.5cm}
\begin{tabular}{|c|c|c||c|c|} \hline
$-t$ (GeV$^2$) & $d\sigma/dt_{ATLAS}$ & $-\Delta_{CNI}$ & $d\sigma/dt_{mod}$ & $-\Delta_{CNI}$  \\   \hline
 0.00029   & 3662 & -33 &  3291  & 417        \\
 0.0004    & 2136 &  33 &  1952  & 296       \\
 0.00051   & 1401 & 146 &  1396  & 229        \\
 0.00067   &  998 & 130 &  1034  & 166           \\
 0.00086   &  797 & 104 &   846  & 130         \\
 0.00112   & 680.1&  75 &   731.1&  98           \\
 0.00143   & 610.6&  62 &  620.7 &  75              \\       \hline
\end{tabular}
\end{table}

\begin{figure}
\includegraphics[width=.6\textwidth]{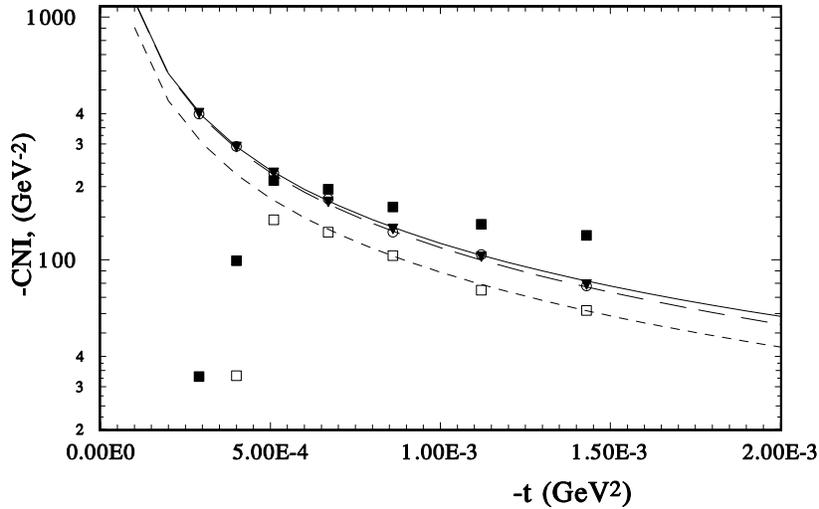}
\vspace{1.5cm}
\caption{  $\Delta_{CNI}(t)$ extracted from experimental data: open squares and full squares -   with different magnitudes of  $\sigma_{tot}(s)=104$ mb and  $\sigma_{tot}(s)=110$ mb with ATLAS form of $F_{h} (t)$; circles - extracted from the model representation of TOTEM data;  hard line - representation of the CNI-term in the form   $F_{Ch}(t)= 8 \alpha_{em} / t $; long dashed line - HEGS model calculations
of the CNI term; short dashed line - model calculations with ATLAS phenomenological fit of $F_{h} (t)$.
  }
\label{Fig_1}
\end{figure}

 The first term proportional to the square of the Coulomb amplitude is calculated exactly and has
 the $1/t^2$ dependence. The second term  proportional to the square of the hadron amplitude
 has small $t$ dependence in the examined $t$ region. The result proportional to the product of Coulomb
 and hadron amplitude has the $1/t$ dependence.   For comparison with the TOTEM data, let us take  our
 model description of all sets at $7 -- 13$ TeV  without  recent  ATLAS data (see below)
  and calculate the differential cross sections at the same points
 of momentum transfer as the ATLAS experimental data at $13 $ TeV.
 For comparison of the $t$ dependence, let us take an artificial term that has a simple $1/t$ dependence
  $F_{Ch}(t)= 8 \alpha_{em} / t $, where $\alpha_{em}$ is a fine electromagnetic structure constant.
  The results are presented in Table 2.   Obviously, the   value of  $\Delta_{CNI}$ (third column  of Table 1)  obtained from the
  ATLAS data has no  normal $1/t$ dependence and differs considerably from the results obtained
  from TOTEM artificial data ( TABLE 2, last column ). In Fig. 1, the data are compared with the CNI term  with hadron amplitudes, used by ATLAS (short dashed line),
   and with HEGS model calculations (long-dashed line), as well as with the term which represents
    the true $1/t$ dependence $F_{Ch}(t)= 8 \alpha_{em} / t $.
     We can see that the model calculations
    exactly reproduce the $1/t$ dependence.
    The points  calculated with the hadron amplitude with $\sigma_{tot}=104$ mb
    are much  lower and give the $1/t$ dependence only for $-t > 0.0005 $ GeV$^2$.
    The points obtained by using    $\sigma_{tot}=110$ mb do not give the $1/t$ dependence for all points.
    It means that  simple additional normalization does not improve the situation.

 In Fig.2, the comparison of the TOTEM and ATLAS data at 13 TeV are presented.
 Obviously, the data of ATLAS above $-t =0.0005$ GeV$^2$ are systematically lower than the TOTEM data. It means that  additional
 corrections of the normalization of  experimental data is required.
 This explains the tension  between the values of $\sigma_{tot}(s)$ presented by the Collaborations.
 In our model fitting, we take into account only statistical errors whereas systematical errors
  are taken into account
 as additional correction coefficients of the normalization.
\begin{figure}
\includegraphics[width=.6\textwidth]{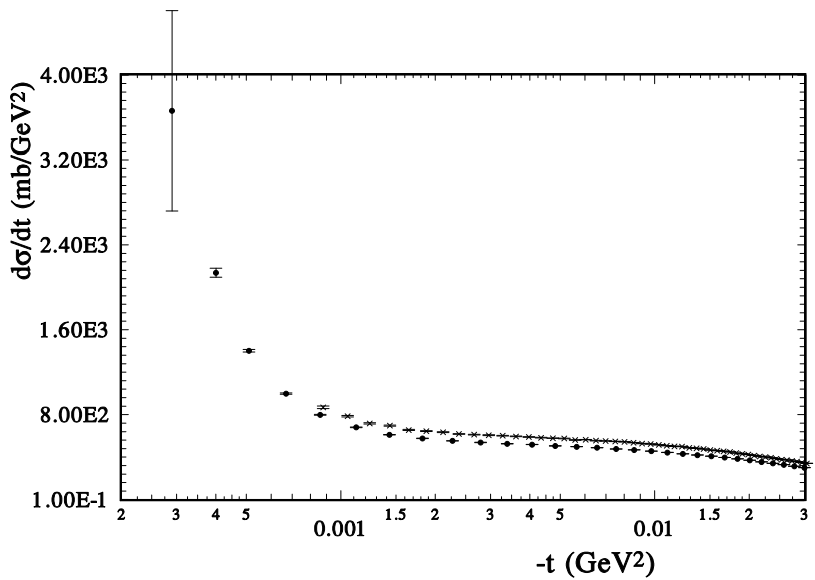}
\vspace{1.5cm}
\caption{  Comparison of experimental data on the differential cross sections at small momentum transfer
of the TOTEM and ATLAS Collaboration
 only  with statistical errors. (Full circles - ATLAS data,
 crosses - TOTEM data.)
  }
\label{Fig_2}
\end{figure}

\section{ Model description of all experimental sets at 7 - 13 TeV}

    There are many different semi-phenomenological models which give a qualitative
    description of the behavior of  differential cross sections of  elastic proton-proton
    scattering at $\sqrt{s} = 13$ TeV (for example \cite{Petrov,Kohara}).
    Some examples can be found in the review \cite{Panch}; hence, we do not give a deep analysis
    of those models.
  Our model was compared with other in \cite{HEGS1}
 (for example, see Table VII).
     One of the common properties of practically all models is that they
    take into account statistic and systematic errors in quadrature forms and in most part
    give only a qualitative description of the behavior of the differential cross section
    in a wide momentum transfer region.
 The analysis of systematic errors, for example, of the TOTEM collaboration,  shows that the  main
  uncertainty is determined by  Luminosity. 
  This affect experimental data
  of one set in one direction.  Additional normalization reflect such a situation.
   This method is used by many other research groups, which  deal with many different sets of experimental data, for example, in analysis of different parton distribution functions (PDF).
In this paper, we use the high energy generalized structure  (HEGS) model
that is based on generalized parton distributions.
   The generalized parton distributions (GPDs) represent the basic properties of the hadron
    structure.
  Different properties of  the hadron structure are reflected in the elastic scattering amplitude.
  Our form of the momentum transfer dependence of  GPDs allows us to
  calculate the electromagnetic form factors, gravitomagnetic form factors, transition form factors,  and Compton form factors.
  As a result, the description of various reactions is based on the same
   representation of the hadron structure.
   This allows us to use
  both hadron form factors (electromagnetic and gravitomagnetic) 
  to build  a high energy generalized structure  model  of elastic nucleon-nucleon scattering.
 The value and the energy and momentum transfer dependence of the real part of the elastic scattering amplitude are determined by the complex energy
     $\hat{s}=s \ exp(-i\pi/2)$. Hence, the model does not introduce some special functions or assumptions
     for the real part of the scattering amplitude.
     The final elastic  hadron scattering amplitude is obtained after unitarization of the  Born term.
    So, first, we have to calculate the eikonal phase
 in the most favorable  eikonal unitarization scheme.
   The eikonal phase corresponds to the Born term of the  scattering amplitude,
   and in the  common case corresponds to
   the spin-dependent potential.

     In  our paper \cite{osc13}, it  was shown that the new data of the TOTEM Collaboration at 13 TeV
     show the existence in the scattering amplitude of  the oscillation term, which can be determined
     by the hadron potential at large distances. In the analysis of experimental data
     of  both the sets of the TOTEM data  additional
     normalization was used. Its values reach sufficiently  large magnitudes. In this case, a very small
     $\chi^2_{dof}$ was obtained with taking into account only statistical errors and
     with  a few number of  free parameters in the scattering
     amplitude, which was obtained in our High Energy Generalized Structure (HEGS) model \cite{HEGS0,HEGS1}.
     However, an additional normalization coefficient reaches a sufficiently large value,
     about $13\%$. It can be in a large momentum transfer region, but is very unusual for a small
     momentum transfer. However, both sets of experimental data (small and large region of $t$)
     overlap in some region and, hence,  affect  each other's normalization.
     It is to be noted   that the value of the normalization coefficient does not impact the magnitudes
     and properties of the oscillation term. We have examined  many different variants of our
     model (including large and unity normalization coefficient), but the parameters of
     the oscillation term have small variations.

     In the present work, 
      all sets of the TOTEM  and ATLAS data at
     $7 -- 13$ TeV (at first, not including recent ATLAS data at 13 TeV)
    are analyzed with  additional normalization equal to unity and taking into account
     only  statistical errors in experimental data.

  To study the new effects in elastic scattering
  as a basis, we take our high energy generalized structure (HEGS) model \cite{HEGS0,HEGS1} which quantitatively  describes, with only a few parameters, the   differential cross section of $pp$ and $p\bar{p}$
  from $\sqrt{s} =9 $ GeV up to $13$ TeV, includes the Coulomb-hadron interference region and the high-$|t|$ region  up to $|t|=15$ GeV$^2$
 and quantitatively well describes the energy dependence of the form of the diffraction minimum \cite{HEGSdm}.
   To determine the new effects, the model is simplified as much as possible.
   To avoid  possible problems
 connected with the low-energy region, we consider here only the asymptotic variant of the model,
  leaving only the hadron spin-flip amplitude, which has  no energy dependence, and  taking it in a simple exponential form with two parameters.
 As was made in all variants of the model, we take into account in the slope
 the $\pi$ meson loop in the $t$-channel, as  was proposed in \cite{Gribov-72,Jenk-72}   
 and recently  studied in \cite{Khoze-Sl00}.
   The total elastic amplitude in general gets five helicity  contributions, but at
   high energy, it is sufficiently to keep only spin-non-flip hadron amplitudes.
   In this case, one take 
    $F(s,t) = F^{h}(s,t)+F^{\rm em}(s,t) e^{\varphi(s,t)} $\, where
 $F^{h}(s,t) $ comes from the strong interactions,
 $F^{\rm em}(s,t) $ from the electromagnetic interactions and
 $\varphi(s,t) $
 is the interference phase factor between the electromagnetic and strong
 interactions \cite{PRD-Sum}.
 Note that all five spiral electromagnetic amplitudes are taken into account
 in the calculation of the differential cross sections.
  It is  essential that in the model only the Born term is taken
   in an analytic form.
 After that, the eikonal phase is calculated by the Fourier-Besel method. Then using the standard
 eikonal representation, the full scattering amplitude is calculated  numerically.
    The Born term of the elastic hadron amplitude at large energy can be written as
    a sum of two pomerons (related with two and three (cross even) gluons)   and
     odderon contributions (cross odd - three gluon amplitude). Note that both pomeron contributions have    the same positive sign, this is an essential difference  from other models.

     \begin{eqnarray}
 F(s,t)_{B}& =& \hat{s}^{\epsilon_0}\left(C_{\mathbb{P}} F_1^2(t)  \ \hat s^{\alpha' \ t} + (C'_{\mathbb{P}} \pm i C'_{\mathbb{O}} t ) A^2(t)  \ \hat s^{\alpha' t\over 4} \right) \; , \\
 \end{eqnarray}
  As seen, all terms have the same intercept  $\alpha_{0}=1+\epsilon_0 = 1.11$, and the pomeron
 slope is fixed at $\alpha^{'}_{0}= 0.24$ GeV$^{-2}$.
 With taking into account the effects of a two $\pi$-meson loop, we have
 $\alpha^{'}=\alpha^{'}_{0} (1+x_{rt}^3 e^{-b_{sl} t Ln(s)})$ with $x_{rt}=\sqrt{4 m_{\pi}-t}$.
  The model takes into account  the electromagnetic  $F_1(t)$ and gravitomagnetic  $A(t)$ form factors,
   which correspond to  the charge and matter
  distributions \cite{GPD-PRD14}. Both form factors are calculated  as the first and second moments of  the same Generalized Parton Distribution (GPDs) function.
  The Born scattering amplitude has  three free parameters (the constants $C_{i}$ ) at high energy:
two for the two pomeron amplitudes  and one for the odderon.
The real part of the hadronic elastic scattering amplitude is determined
   through the complexification $\hat{s}=-i s$ to satisfy the dispersion relations.
   The oscillatory function was determined \cite{osc13}
\vspace{-0.1 cm}
\ba
  f_{osc}(t)=h_{osc} (1+i) J_{1}(\tau))/\tau;  \ \tau = \pi \ (-t)/t_{0} \ G^{2}_{em}(t),
\ea
here $J_{1}(\tau)$ is the Bessel function of the first order
and $ G_{em}(t)$ - the electromagnetic form factor of the proton.
 This form has only  two additional fitting parameters and allows one to represent
 a wide range of  possible oscillation functions.

\begin{figure}
\includegraphics[width=.45\textwidth]{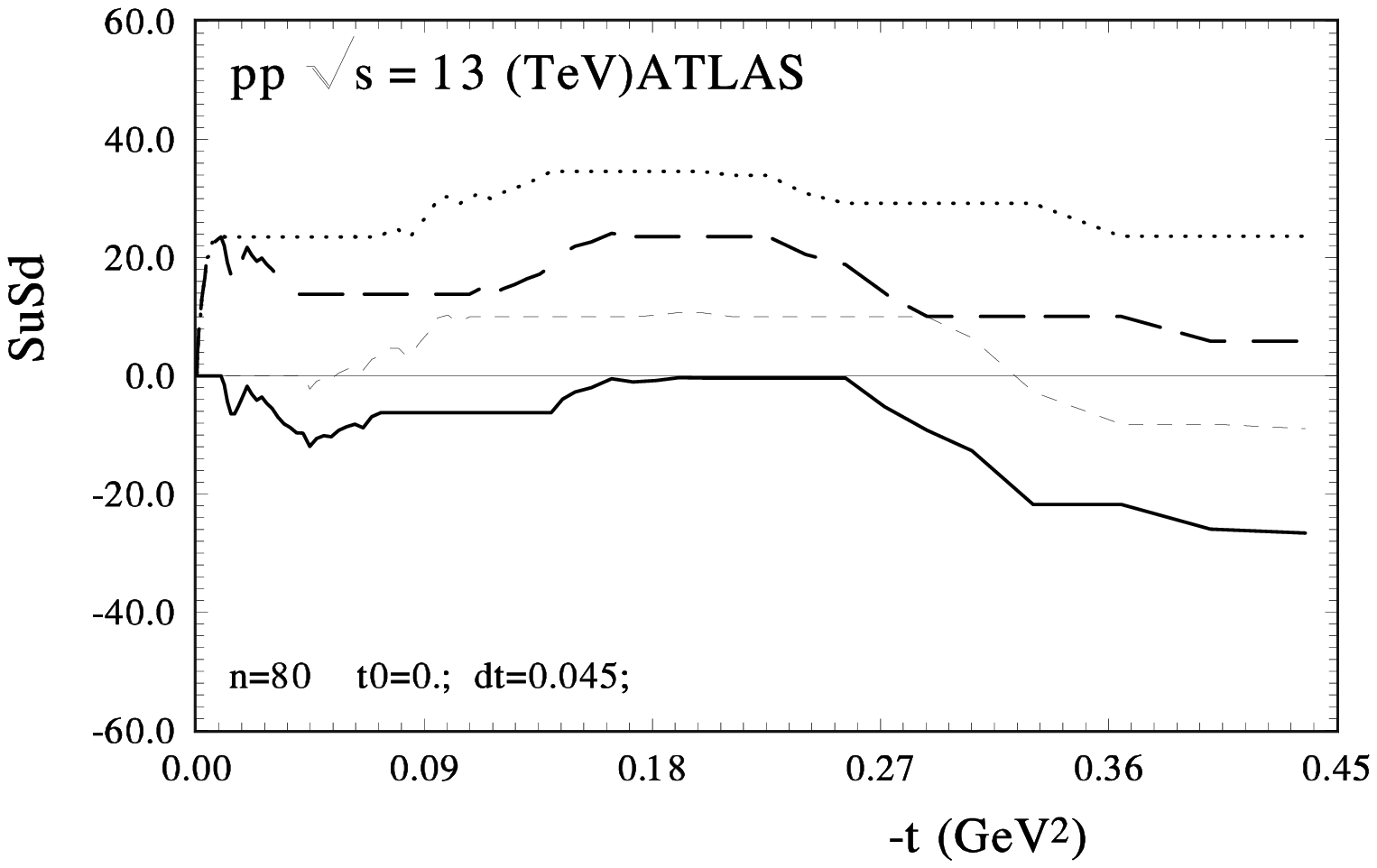}
\includegraphics[width=.45\textwidth]{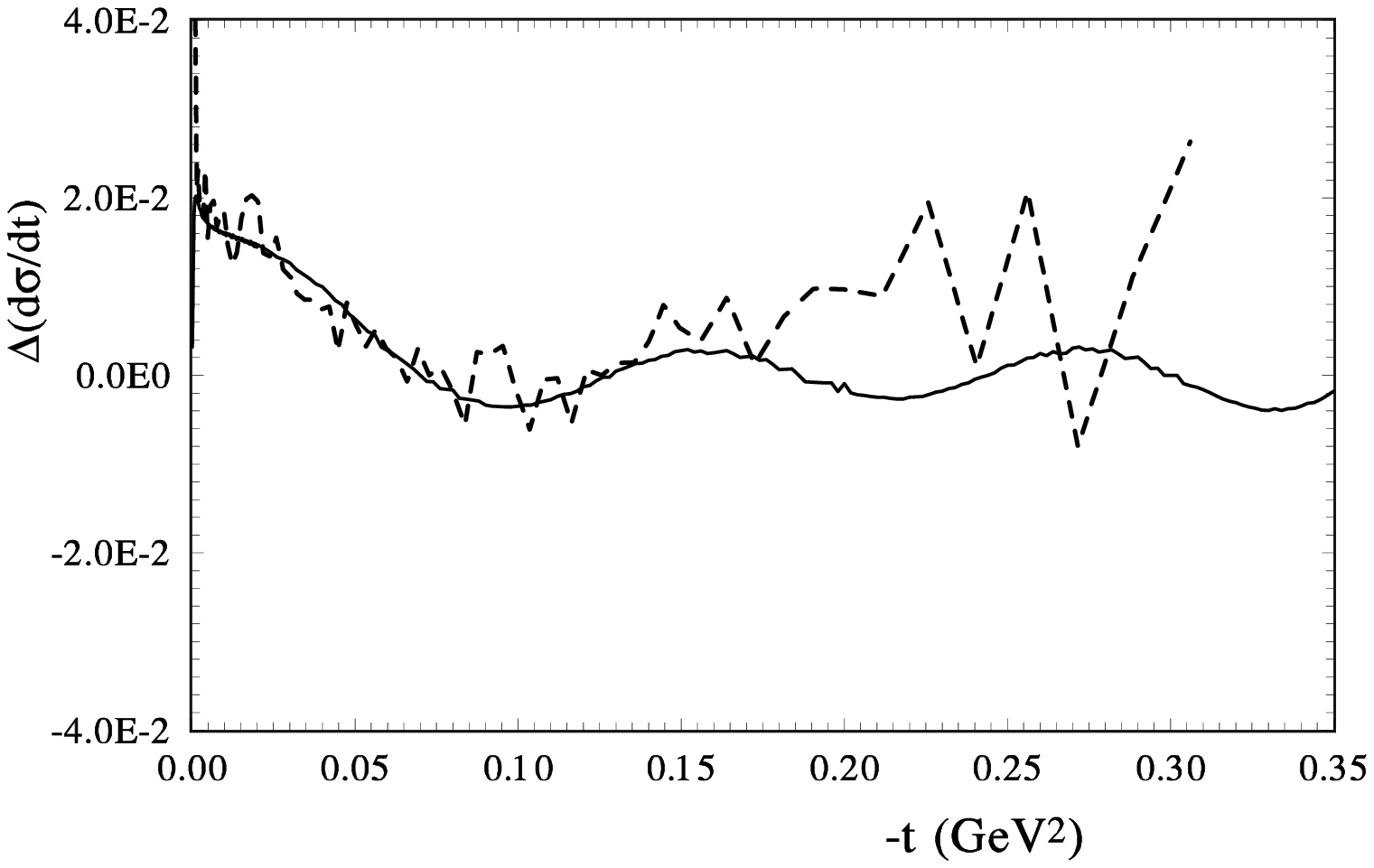}
\vspace{1.5cm}
\caption{ a) (left side) The effect of the oscillation term in the even and odd parts of
the intervals of momentum transfer of ATLAS data at $13$ TeV (hard and dotts lines)
The same but with shifted  intervals by half the period - (short dashed lines);
b)  (right side)  $R \Delta_{th}$  of eq.(6) (the hard line)
and $R \Delta_{ex.} $ eq.(7) (the tiny line) of ATLAS data at $13$ TeV.
  }
\label{Fig_3}
\end{figure}

 As was found in \cite{fd13}, 
 the behavior of experimental data requires an additional term
 to scattering amplitude in the CNI region, which is determined by the hadron interaction at large distances. In the present work, it  takes
 a simple exponential form
  \ba
 F_{d}(t)=h_{d} (i-\rho(t))  e^{- (B_{1} |t| + B_{2} |t|^2) \log{\hat{s_{13}} }} \ G^{2}_{em}(t),
 \label{fd-exp}
\ea
where  $\rho(t)$ is determined by the main amplitude $F(s,t)$.
  Taking into account all sets of  experimental data (Table 1),
  715 experimental points were used  in the fitting procedure with thirteen free parameters.
  After the fitting procedure, we obtain  $\chi^2/n.d.f. =1.06$ (remember that we used only statistical errors) with additional normalization coefficients
 $$ n_{A1}=0.97, n_{T1}=1.02, n_{T2}=0.96, n_{A2}=0.96, n_{T3}=1.07, $$
 $$ n_{T4}=0.96, n_{T5}=1.09, n_{T6}=1.09, n_{A3}=0.96$$
If we add the experimental data of the TOTEM Collaboration at $\sqrt{s}=2.76$ TeV, the number of experimental points
 increases to $778$ and we obtain $\chi^{2}_{dof} =1.07$.

\begin{figure}
%
\includegraphics[width=.45\textwidth]{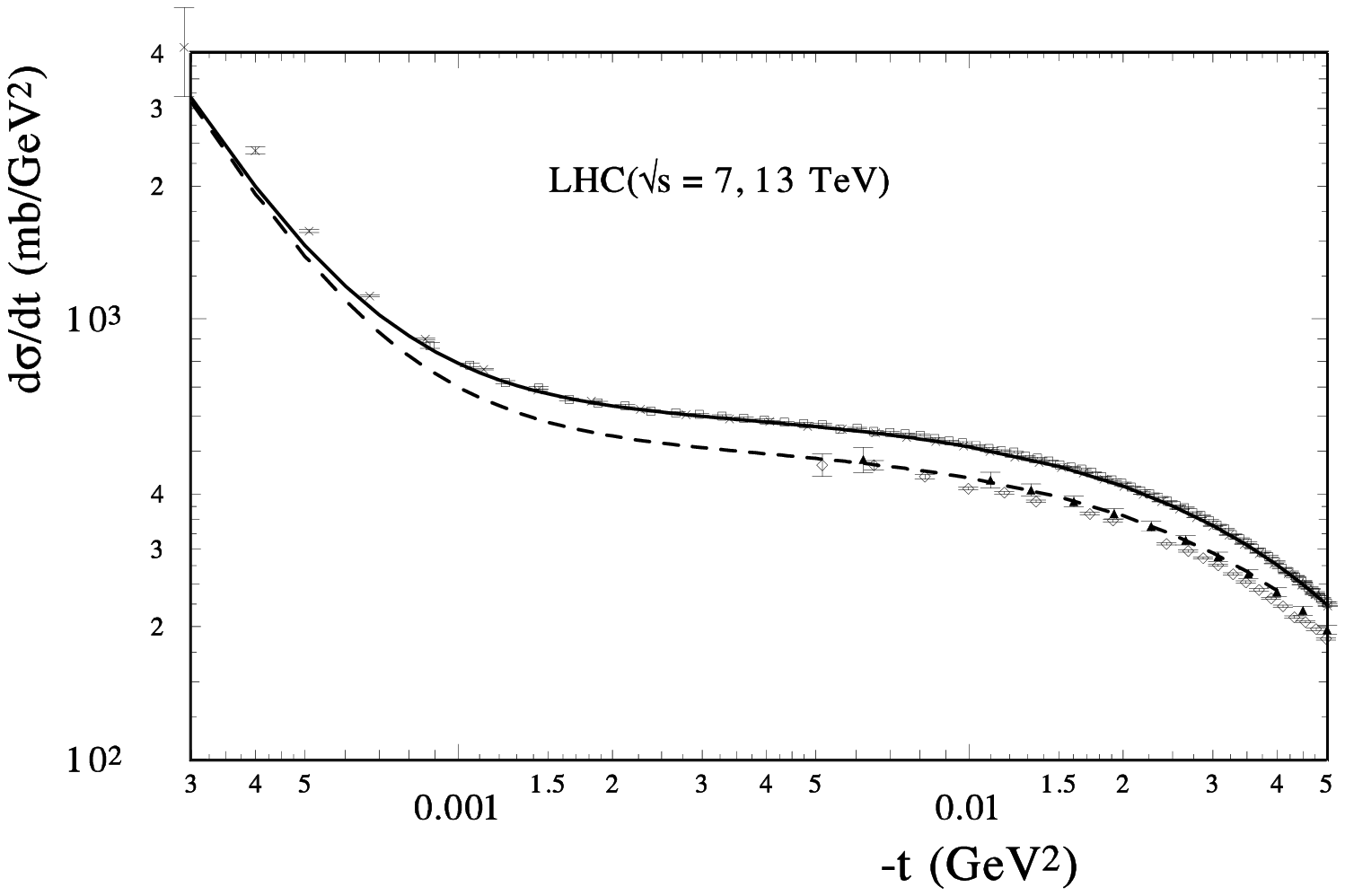}
\includegraphics[width=0.45\textwidth]{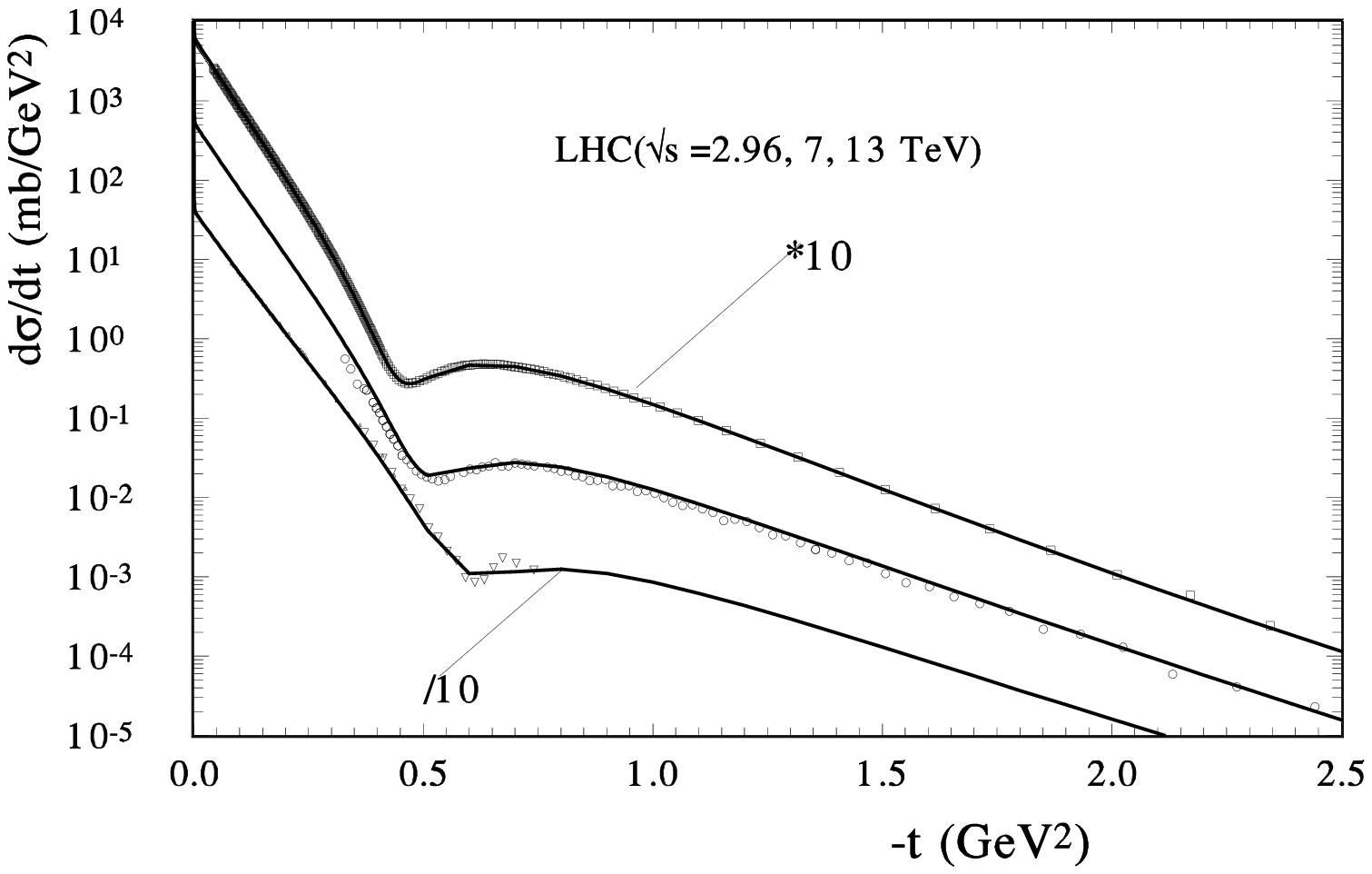}
\vspace{1.5cm}
\caption{The differential cross sections are calculated in the framework of the HEGS model
 at $2-13$ TeV (the experimental points with only statistical errors).
  a) [right] the full region of $t$; 
  b) [left] the magnification of the region of the small momentum transfer;
 (the data of ATLAS ($\sqrt{s}=13$ TeV) are drawn without an additional coefficient of the normalization }.
\label{Fig_4}
\end{figure}

  The constants of the oscillatory and anomalous terms are determined with high accuracy
  $$h_{osc}=0.36\pm 0.013; \ \ \ h_{d} = 1.41 \pm 0.07 GeV^{-2}.$$
  They can be compared with the same constants obtained from the
   analysis of the  TOTEM data at 13 TeV
  \cite{osc13,fd13}
   $$h_{osc}=0.16\pm 0.02; \ \ \ h_{d} = 1.7 \pm 0.01 GeV^{-2}.$$
  It can be seen that additional experimental information leads to an increase in the magnitudes of the oscillation term. If such peculiarities are related with the specific properties of  the TOTEM 13 TeV data only, the constants have to decrease and
  errors increase essentially.
  Note that the form and value of the oscillation are independent of  additional normalization of the set of experimental data. The normalization shifts all data of one set in one direction. The oscillation term  lead to increasing or
     decreasing some parts of the model calculate differential cross sections.
     Hence, it is reflecting the deviations from the average position of experimental data of the set.

\begin{figure}[b]
%
\includegraphics[width=.45\textwidth]{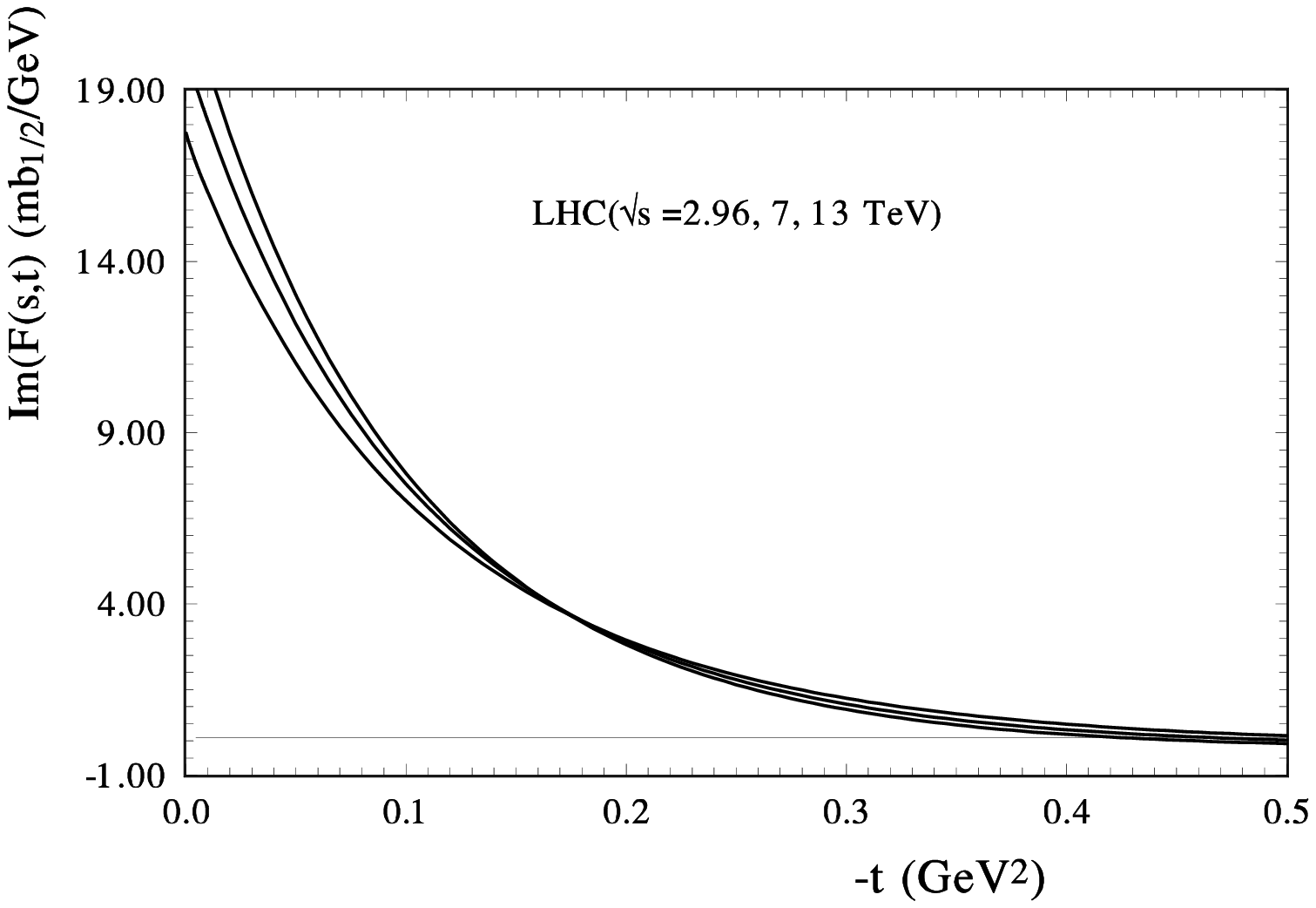}
\includegraphics[width=0.45\textwidth]{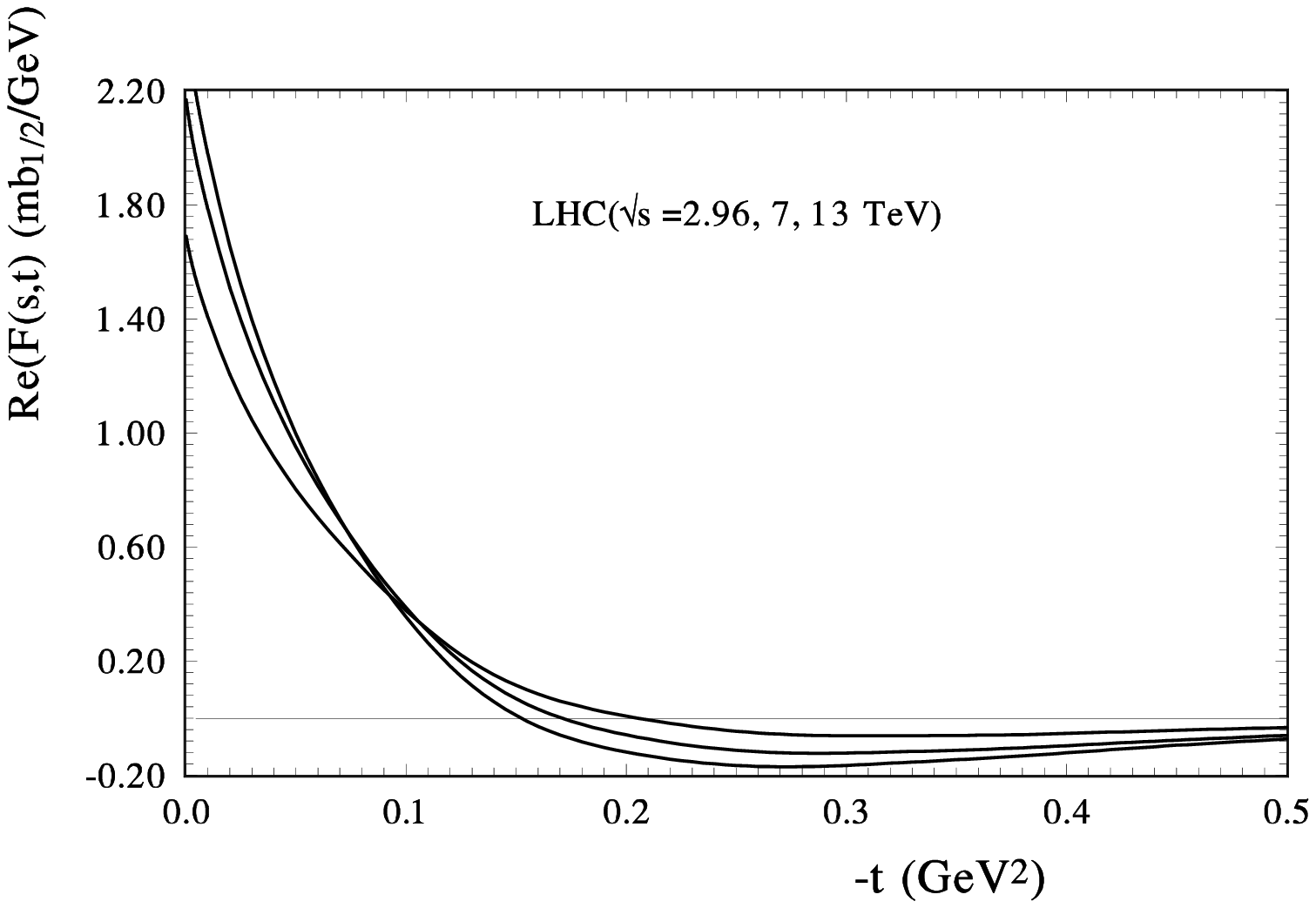}
\vspace{1.5cm}
\caption{ The  $s$ and $t$ dependence of the imaginary (left)and real (right) parts of
the elastic scattering amplitude (lines  at small $t$ correspond top down
 to 13, 7, 2.96 TeV).
  }
\label{Fig_5}
\end{figure}

  Now let us add the new data of the ATLAS Collaboration at $13$ TeV into our consideration.
  As there are some problems with the points in the CNI region, let us take them with some increasing errors
  (for example,  twice as large or  proportional to the error of the first point).
  The number of experimental points is 795 and  $\chi^2 = 903$
  is obtained.
  In this case, additional normalization for  the  ATLAS data is $n_{A3}=1.06$,
  and  the normalization of the  TOTEM data at $13$ TeV is $n_{T6}=0.94$.
  This reflects the tension between these data.
  But the magnitudes of the corresponding constants of the new terms practically do not change.
    This means that such additional information does not change the determination of these peculiarities.
    Taking into account the experimental data of the TOTEM Collaboration
    at $2.9$ TeV   leads  to the same result.   In this case,
     the number of experimental points increases to $858$
    and $\chi^2=984$ is obtained. But again, the magnitudes of the constants practically do not change.

   The contributions of the oscillatory term in the differential cross section of the ATLAS data at $13$ TeV    are presented in Fig.3 (a,b). In Fig.3a, the results of the statistical method of the two    chosen independent selections are shown.
    This is the result of the procedure in which
    the whole examined region of momentum transfer is divided into some equal intervals, and then deviations of experimental data from some smooth  curve
   in units of errors of experimental data were summed separately in the odd and even intervals.
   If the some periodic structure is present in experimental data, it gives increasing sums of the
   contribution in the odd and even intervals in the case of coincidence of the period  with the span of the interval. If the beginning of intervals is shifted by half the period, the sums will be near each other.

 The results of the description of the differential cross sections are represented in Fig.4(a,b).
   We can see that the diffraction minimum has a sharp form at $7$ and $13$ TeV, but a less pronounced at $\sqrt{s}=2.79$ TeV.

 In Fig.5, the  $s$ and $t$ dependence of the imaginary (left)and real (right) parts of
the elastic scattering amplitude obtained in the model with taking into account  additional effects are represented. Note that the real part of the scattering amplitude has the first zero at small momentum transfer
 which changes with energy. The imaginary part of the scattering amplitude has the first zero at
 the essentially large momentum transfer. It determines the position of the diffraction minimum.
 Note that there is a crossing point on which the curves intersect. It is determined by the growth of the total cross sections and, simultaneously, the growth of the slope of the scattering amplitude.

\section{ Results and discussion }
 Now we have some tensions between the results of the TOTEM  and ATLAS Collaborations at  LHC all energies.
  This  especially concerns the extraction of magnitudes of
the total cross sections and $\rho (t=0)$. New unique results of the ATLAS Collaboration
do not improve the situation. The new  ATLAS data present for the first time the results
 at very small momentum transfer.
 These first points and common difference between the ATLAS and TOTEM
 data will be crucial for the discussion of the total cross
 sections, rho-parameter, Odderon contributions, et cetera.
 Unfortunately, we found out that  these points
 contradict the standard behavior of the electromagnetic amplitude.
  The analysis shows that the difference between the other points of ATLAS and TOTEM is
 related not with the behavior of the scattering amplitude
 but with different normalization only, which is very important for other studies.

    Examining the new peculiarity in the differential cross sections of  elastic scattering
    at LHC energies in a wide momentum transfer  on the basis of all experimental data  of the TOTEM-CMS and  AlFA-ATLAS Collaborations, previously found in the differential cross section of the TOTEM Collaboration at 13 TeV \cite{osc13,fd13}, strongly confirm the existence of these effects.
    Especially, we take only statistical errors which significantly reduce the corridor of the possible behavior of the   scattering amplitude. The systematical errors are taken into account as additional normalization,
     which are equal  for all points of the set.

     The magnitude of the constants of the effects is relatively large and is determined  with small errors ($h_{fd}=1.54 \pm 0.08$ GeV$^{-2}$ and $h_{osc}=0.37 \pm 0.014$ GeV$^{-2}$),
    depending on the new ATLAS data at 13 TeV, and
     ($h_{fd}=1.41 \pm 0.07$ GeV$^{-2}$ and $h_{osc}=0.36 \pm 0.013$ GeV$^{-2}$),
      and taking into account the new ATLAS data at 13 TeV.

\begin{table*}[tbp]
\label{Table_3}
\caption{  The  $\sigma_{tot}(s)$ and $\rho(s)$ obtained in the HEGSh model}
\vspace{.5cm}
\begin{tabular}{|c|c|c|c||c|c|c|} \hline
 $\sqrt{s}$ (GeV)   & $\sigma_{tot-ATLAS}$& $\sigma_{tot-TOTEM}$  & $\sigma_{tot-model}$
 & $\rho_{tot-ATLAS}$& $\rho_{tot-TOTEM}$  & $\rho_{tot-model}$  \\ \hline
  7.  &$95.35 \pm 2. $ &  $98.5  \pm 2.9 $ & $98.8 \pm 1.4$     &   &  $0.145\pm0.09$ &$0.117 \pm 0.03$ \\
  8.  &$102.9 \pm 2.3$ &  $101.7 \pm 2.9 $ & $100.7 \pm 1.4$ &  & $0.12\pm0.03$  &$0.118 \pm 0.04$ \\
  13. &$104.7 \pm 1.1$ &  $110.6 \pm 0.6 $ & $108.1 \pm 1.4$ &$0.098 \pm 0.011$   &$0.1 \pm 0.01$  &$0.119 \pm 0.04$  \\  \hline
\end{tabular}
\end{table*}
%

The HEGS model describes at a  quantitative level the new experimental data at $2 - 13$ TeV
$\chi^{2}_{dof}=1.09$ without the new ATLAS data and   $\chi^{2}_{dof}=1.16$ with
taking into account only statistical errors (see Fig.5).
Such a description can be obtained only considering
the new peculiarities of the elastic differential cross sections.

The new phenomena are determined by  hadron interactions at large distances.
The  corresponding  hadron potential has no simple exponential behavior.
Our calculations  \cite{Conf-22} show that it has a  modified Gaussian form with probably a cut at large distances
$V(r)=h Exp[-B r^2]/(R_{0}+r)$.
This may be due to glueball states of the gluon.
In such a form, the gluon can be distributed at large
distances above the confinement level.

The new effects can impact  the determination of values
of the total cross sections, the ratio of
the elastic to the total cross sections, and
$\rho(s,t)$, the ratio of the real to imaginary part of the elastic scattering
amplitude.
As a result, we have obtained the values of $\sigma_{tot}(s)$ and $\rho(s)$
represented in Table 3.

It is very likely that such effects exist also in  experimental data
at essentially lower energies \cite{osc-conf},
but, maybe, they have a more complicated form
(with two different periods, for example).

\section{Conclusion}


Our critical analysis of the new experimental data obtained by
		the ATLAS Collaboration group at 13 TeV shows that the first experimental points
		contradict the behavior of the electromagnetic amplitude at small momentum
		transfer. To remove such contradiction, it is necessary to take these points with  large errors.
		The problem of  tension
		between the ATLAS and TOTEM Collaborations data lies more part in different normalizations.
		
		The analysis is based  on all sets of experimental data of the elastic
		differential cross section in a wide region of momentum transfer confirms
		the existence of the
		new effects discovered on the basis of experimental  data at 13 TeV \cite{osc13,fd13}
		and associated with the specific properties of the  hadron potential
		at large distances.
As a result, we have shown that the new peculiarities in the scattering
amplitude are confirmed by all LHC  data and are not artificial or instrumental effects of one experiment.
The existence of such peculiarities is important for our understanding of
the hadron interaction at large  distances. Hence, it opens  up a new direction
   in the study of hadron structure and interaction at super high energies and large distances.

\vspace{2.cm}

\end{document}